\documentclass{aastex63}
\received{..., 2022}
\revised{..., 2022}
\accepted{\today}
\submitjournal{ApJ}

\shorttitle{TSI during the last five centuries}
\shortauthors{Penza et al.}
\graphicspath{{./}{figures/}}

\begin{document}
\title{Total Solar Irradiance during the Last Five Centuries}

\correspondingauthor{Francesco Berrilli}
\email{francesco.berrilli@roma2.infn.it}

\author[0000-0002-3948-2268]{Valentina Penza}
\affiliation{Dipartimento di Fisica, Universit\`a di Roma Tor Vergata, Via della Ricerca Scientifica 1, Roma, 00133, Italy}

\author[0000-0002-2276-3733]{Francesco Berrilli}
\affiliation{Dipartimento di Fisica, Universit\`a di Roma Tor Vergata, Via della Ricerca Scientifica 1, Roma, 00133, Italy}

\author[0000-0002-1155-7141]{Luca Bertello}
\affil{National Solar Observatory, 3665 Discovery Dr., Boulder, CO 80303, USA}

\author[0000-0003-4898-2683]{Matteo Cantoresi}
\affiliation{Dipartimento di Fisica, Universit\`a di Roma Tor Vergata, Via della Ricerca Scientifica 1, Roma, 00133, Italy}

\author[0000-0002-4525-9038]{Serena Criscuoli}
\affil{National Solar Observatory, 3665 Discovery Dr., Boulder, CO 80303, USA}

\author[0000-0001-5625-9781]{Piermarco Giobbi}
\affiliation{Dipartimento di Fisica, Universit\`a di Roma Tor Vergata, Via della Ricerca Scientifica 1, Roma, 00133, Italy}




\begin{abstract}
The total solar irradiance (TSI) varies on timescales of minute to centuries. On short timescales it varies due to the superposition of intensity fluctuations produced by turbulent convection and acoustic oscillations. On longer scale times, it changes due to photospheric magnetic activity, mainly because of the facular brightenings and dimmings caused by sunspots. While modern TSI variations have been monitored from space since 1970s, TSI variations over much longer periods can only be estimated using either historical observations of magnetic features, possibly supported by flux transport models, or from the measurements of the cosmogenic isotope (e.g., \textsuperscript{14}C or \textsuperscript{10}Be) concentrations in tree rings and ice cores. The reconstruction of the TSI in the last few centuries, particularly in the 17th/18th centuries during the Maunder minimum, is of primary importance for studying climatic effects. To separate the temporal components of the irradiance variations, specifically the magnetic cycle from secular variability, we decomposed the signals associated with historical observations of magnetic features and the solar modulation potential $\Phi$ by applying an Empirical Mode Decomposition algorithm. Thus, the reconstruction is empirical and does not require any feature contrast or field transport model. The assessed difference between the mean value during the Maunder minimum and the present value is $\simeq2.5 Wm^{-2}$. Moreover it shows, in the first half of the last century, a growth of $\simeq 1.5 W m^{-2}$ which stops around the middle of the century to remain constant for the next 50 years, apart from the modulation due to the solar cycle.
\end{abstract}
\section{Introduction} 
\label{sec:intro}
Solar Irradiance is the Earth's primary energy input \citep{kren2017} and is a fundamental ingredient for the understanding and characterization of a large variety of phenomena, which include the modeling of terrestrial global or regional climate variations
\citep[e.g.][]{haigh2010,Lockwoodetal2010, lean2017, jungclaus2017,Schmutz2021}, ecosystems \citep{hader2011},  telecommunications and renewable energies applications \citep{myers2017}. Recently, a surge in studies of solar irradiance and its variability has been driven by the increasing interest in understanding and modeling the irradiance variability in other stars \citep{fabbian2017,faurobert2019, kopp2021}, which, likewise for the Earth, affects exoplanets atmospheres and it is fundamental to assess the climate and habitability of planets orbiting solar-like stars \citep[e.g.][]{linsky2019,Galuzzo2021}. \\
Solar Irradiance is the electromagnetic energy emitted by the Sun in the unit of time and area that fall outside the earth's atmosphere at a distance of one Astronomical Unit. The term Total Solar Irradiance, or TSI, refers to the irradiance integrated over the disk and over the whole energy spectrum.
Precise measurements of the TSI have only existed since 1978, as they require the ability to perform observations out of the Earth atmosphere, which absorbs large fractions of the radiation.
These observations showed that the TSI varies over time scales from minutes to years and decades \citep[e.g.][]{woodard1983, kopp2016} 
and that variations from days to decades are clearly modulated by the solar surface magnetism \citep{wilson1978, hudson1988}. 
Variations over temporal scales longer than approximately the 11-years solar cycle are difficult to assess, as space missions rarely extend over more than a decade and therefore the creation of long-term irradiance dataset rely on precise intercalibration of measurements obtained with different instruments.\\
Spurred by the necessity of explaining measured variations on one hand, and producing long records essential to quantify the effects of solar variability on the Earth climate, several models have been developed to reproduce the measured TSI and estimate its variability in the past, before measurements started to be available \citep[e.g.][]{frohlich2004, coddington2016, tapping2007, yeo2014, wu2018}. Although often based on very different approaches, all these models take into account mainly the combinations of the effects due to active regions, namely dark sunspots and bright regions network and facular area. It is well known that in this sort of competition, the emission of bright ARs on average exceeds the sunspot deficit, causing a net increase in total irradiance in phase with the magnetic activity level, with a difference between the maximum and minimum of nearly 0.1 \%.
Models of solar irradiance generally fall into one of two categories: proxy based or semi-empirical.\\ 
Proxy models reproduce irradiance variability by regressing irradiance measurements with proxies of magnetic activity. For example, both the Naval Research Laboratory model \citep[NRL,][]{coddington2016, lean2020} and the EMPirical Irradiance REconstruction \citep[EMPIRE,][]{2017JGRA..122.3888Y} make use of sunspot area and MgII core-to-wing ratio regressed on TSI and SSI measurements to take into account of the sunspot and plage contributions, respectively.
In this context, the attempts to reconstruct solar variability also in regions of the spectrum, i.e., ultraviolet, are particularly relevant for the energetic balancing of the earth's upper atmosphere. \citep[e.g.][]{Kutiev2013, Bordi2015,lovric2017, Criscuoli2018, Rodriguez2019, Bertello2020, Berrilli2020}.\\
Semi-empirical models, often referred to as physics-based, combine area coverage of quiet and magnetic features, typically derived from full-disk observations,  with their corresponding radiative emission. Typically,  chromospheric full-disk daily observations, e.g. CaII K index,  or coronal observations,  are employed in combination with Non Local Thermodynamic Equilibrium (Non-LTE) spectral syntheses obtained with sets of semi-empirically derived atmospheres \citep[e.g. FAL models][]{fontenla1999,fontenla2011, haberreiter2011,fontenla2015} to reproduce both the Total \citep[e.g.][]{ermolli2011, fontenla2011} and the Spectral Solar Irradiance \citep{penza2003,fontenla2011,haberreiter2014,Fontenla2017, Criscuoli2018,criscuoli2019}. The Spectral and Total Irradiance Reconstructions \citep[SATIRE,][]{krivova2007, ball2014} makes use of full-disk daily magnetograms to single out magnetic and quiet regions, to which are associated radiative emissions computed under LTE making use of Kurucz and modified FAL models.  Typically, irradiance reconstruction models reproduce more than 90\% of the observed TSI variability observed at the solar-cycle time scale \citep[e.g.][]{lean2020}.\\
Variations over longer temporal scales and their physical origins are still debated. On one hand, as mentioned above assessing these variations through measurements is extremely difficult \citep[e.g.][]{kopp2014,dewitte2017,dudokdewit2020}, on the other, different irradiance reconstruction models produce different levels of variability \citep[e.g.][]{2017JGRA..122.3888Y,lean2020}. From a theoretical perspective, although different mechanisms have been suggested \citep[see for instance the recent reviews by][]{faurobert2019,petrie2021}, the most accepted cause for long-term irradiance variability are variations in the properties of the quiet Sun \citep[e.g.][]{foukal2011, lockwood2020}. This hypothesis has been recently corroborated by \citet{rempel2020}, who, making use of state-of-the-art magneto hydrodynamic simulations of the solar photosphere,  showed that variations as small as 10\% of the quiet Sun magnetic properties may induce changes of the TSI amplitude comparable to the 11-years cycle variations. Understanding the properties of the quiet Sun magnetic field and its variations over the solar cycle and longer temporal scales is an active area of research \citep[e.g.][]{schnerr2011, criscuolifoukal2017,faurobert2020} and is part of the Critical Science Plan \citep{rast2021} of the NSF operated Daniel K. Inouye Solar Telescope \citep{rimmele2020}.\\
The contribution of quiet Sun to solar irradiance is higher, relatively speaking, during periods of minima. It is known that the solar magnetic activity has crossed grand minima periods, such as the Maunder minimum during the years of 1645 to 1715. 
Estimate the average level of TSI
during these period of minima, and compare it with present measured values, is essential to model natural contributions to the increase of regional or global temperatures observed from the pre-industrial era \citep[e.g.][]{Shindell2001,Inesonetal2015,jungclaus2017, matthes2017}. Models have produced a wide range of possible TSI scenarios for the Maunder minimum, ranging from being comparable \citep[e.g.][]{Schrijver2011} to as much as 0.35\% \citep[e.g.][]{egorova2018} of the current day quiet Sun. The value adopted by the Paleoclimat Modeling Intercomparison Project-4 is 0.055\% \citep{jungclaus2017}, which falls in between  the 0.04\% and the 0.067\% estimated more recently with SATIRE-H \citep{wu2018} and  NRLTSI-2 \citep{lean2018} models, respectively.
It is important to note that irradiance reconstructions prior to the twentieth century are necessarily based on the use of proxies, full-disk observations of the Sun being available only starting at the end of the nineteenth century. Two proxies are typically available: sunspot numbers (or groups), dating back to 1610, and radioisotopes, which allow to estimate the solar activity over millennia \citep{usoskin2017,brehm2021}. The models range from using correlation  between irradiance and proxies derived from modern observations \citep{steinhilber2009, lean2018}, to the use of models of various degree of complexity to derive the distribution of the magnetic fields over the disk  \citep[e.g.][]{wang2005,bolduc2012,wu2018, Berrilli2020}.\\
In this work we present a novel method to estimate the mean levels of TSI, averaged at 22-y and useful for global or regional climatology studies, and to hypothesize a possible variability of solar magnetic structures, especially sunspots and plage, on a shorter scale, i.e., one year. This method
relies on the use of an Empirical Mode Decomposition (EMD) algorithm to separate the different temporal components of the solar activity variability present in the signal  
of the solar modulation potential $\Phi$ to estimate the contribution of sunspot, faculae and quiet magnetic fields to irradiance variability.
The Solar modulation potential $\Phi$ is defined as the mean energy loss per unit charge by the high-energy charged particles forming the Galactic Cosmic Rays as they propagate through the heliosphere. As discussed in the next section, $\Phi$ is modulated by solar activity.
EMD is an adaptive and effective algorithm to deal with nonlinear, nonstationary signals and to identify quasi-periodic embedded structures.

\section{Dataset and method} \label{sec:sec1}
The models that we developed is based on the widely common accepted assumption that on scales from days to the 11-years cycle the TSI is modulated by the positive contribution of faculae \citep[e.g.][]{Pagaran2009} and network \citep[e.g.][]{Berrilli1999,ermolliI2003} and the negative contribution of sunspots \citep[e.g.][]{wenzler,ball2012}. Moreover, we assume that variations of the quiet Sun magnetism modulate irradiance at longer temporal scales and that such variations are independent from global dynamo processes.  The last assumption is that variations of the solar modulation potential are proxies for the evolution of the solar surface magnetism. This assumption is based on the evidence that solar activity modulates galactic cosmic rays (GCRs) that enter the heliosphere \citep[e.g.][]{Martucci2018} and that the GCRs modulate some isotopes present in the Earth's atmosphere, e.g., \textsuperscript{14}C 
or \textsuperscript{10}Be.
All the  assumptions employed in our model are common to other models described in Sec. \ref{sec:intro}. However, our method differ in the approach employed to estimate the contributions of the different components to the TSI variability from the solar modulation potential. The method, described in detail in the next sections, is briefly summarized here.\\
We adopted two main composite data sets based on measurements of plages and sunspot areas. Both data sets are publicly available from the Max Plank Institute (MPI) website (http://www2.mps.mpg.de/projects/sun-climate/data.html).
The first composite consists of values derived from Ca II K spectroheliogram observations covering the period 1892 to 2019 \citep{chatzistergos19}. The second composite includes measurements of sunspot group area, daily total sunspot area and the photometric sunspot index (PSI) for the period 1874-2019, calculated after cross-calibration of measurements by different observers \citep{mandal2020}.\\ Plage and sunspot areas at times when measurements were not available were estimated using the method described in Sec.~\ref{sec:sec3}, which makes use of the correlations between the plage and sunspot areas and the solar modulation potential. 
To this aim, we employ annual estimates of the solar modulation potential  derived from the analysis of the \textsuperscript{14}C \citep{muscheler2007} \footnote{The dataset is available on the NOAA website (https://www.ncdc.noaa.gov/paleo-search/).} extending from 1000 to 2001 A.D.\\
The Empirical Mode Decomposition \citep[e.g.][]{1998RSPSA.454..903H} of the solar modulation potential, described in Sec.~\ref{sec:sec4}, provides the long term modulation, while the dimensionless weights of the plage and sunspot coverages are obtained by fit with available TSI data, as provided by PMOD composite \citep{willson14} \footnote{available at https://www.pmodwrc.ch/en/research-development/solar-physics/tsi-composite/}.\\
\section{Plage and sunspot coverages reconstruction} \label{sec:sec3}
\begin{figure}[htbp]
\centering
\includegraphics[width=0.47\linewidth,trim=4cm 0cm 4cm 0cm,clip]{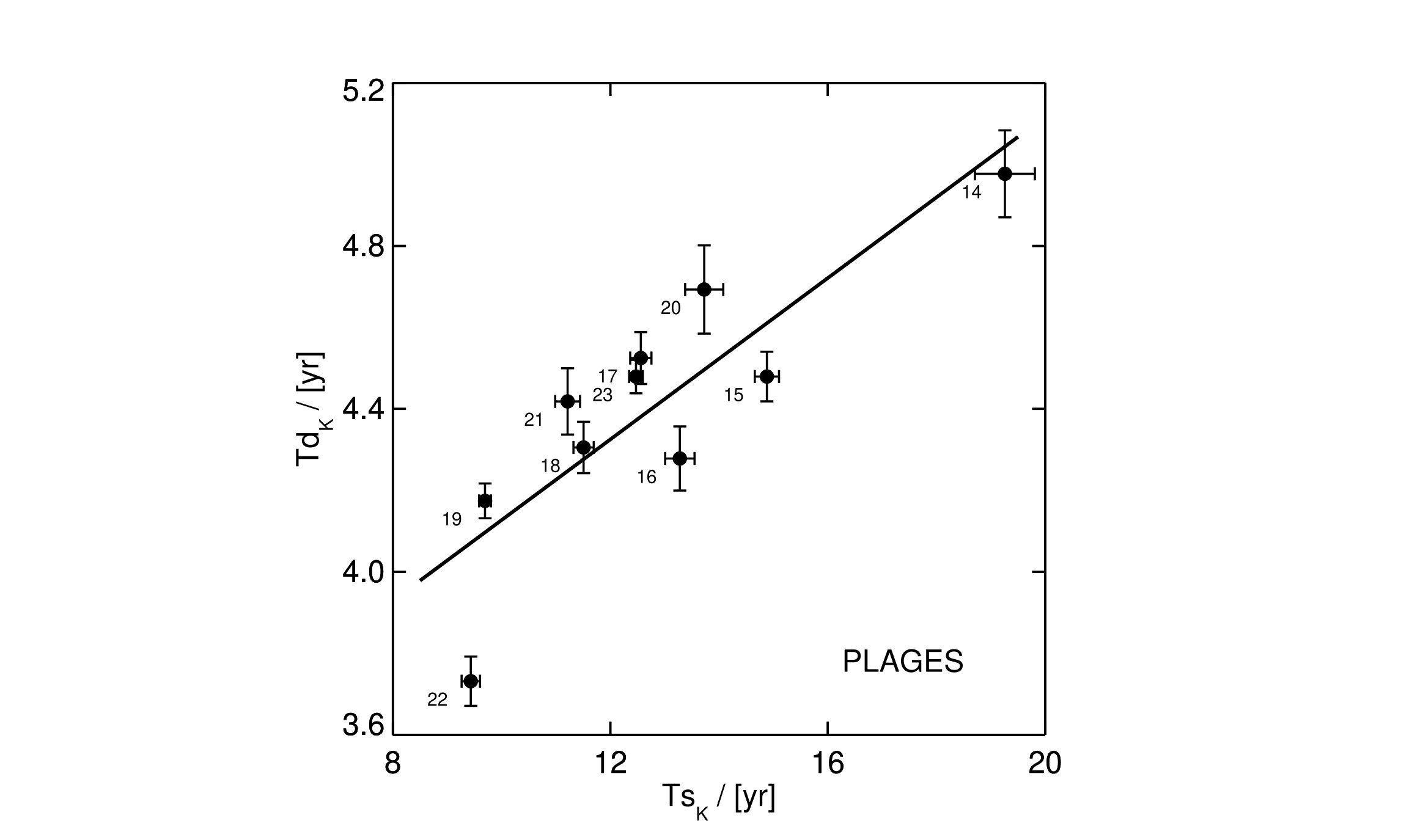}
\includegraphics[width=0.47\linewidth,trim=4cm 0cm 4cm 0cm,clip]{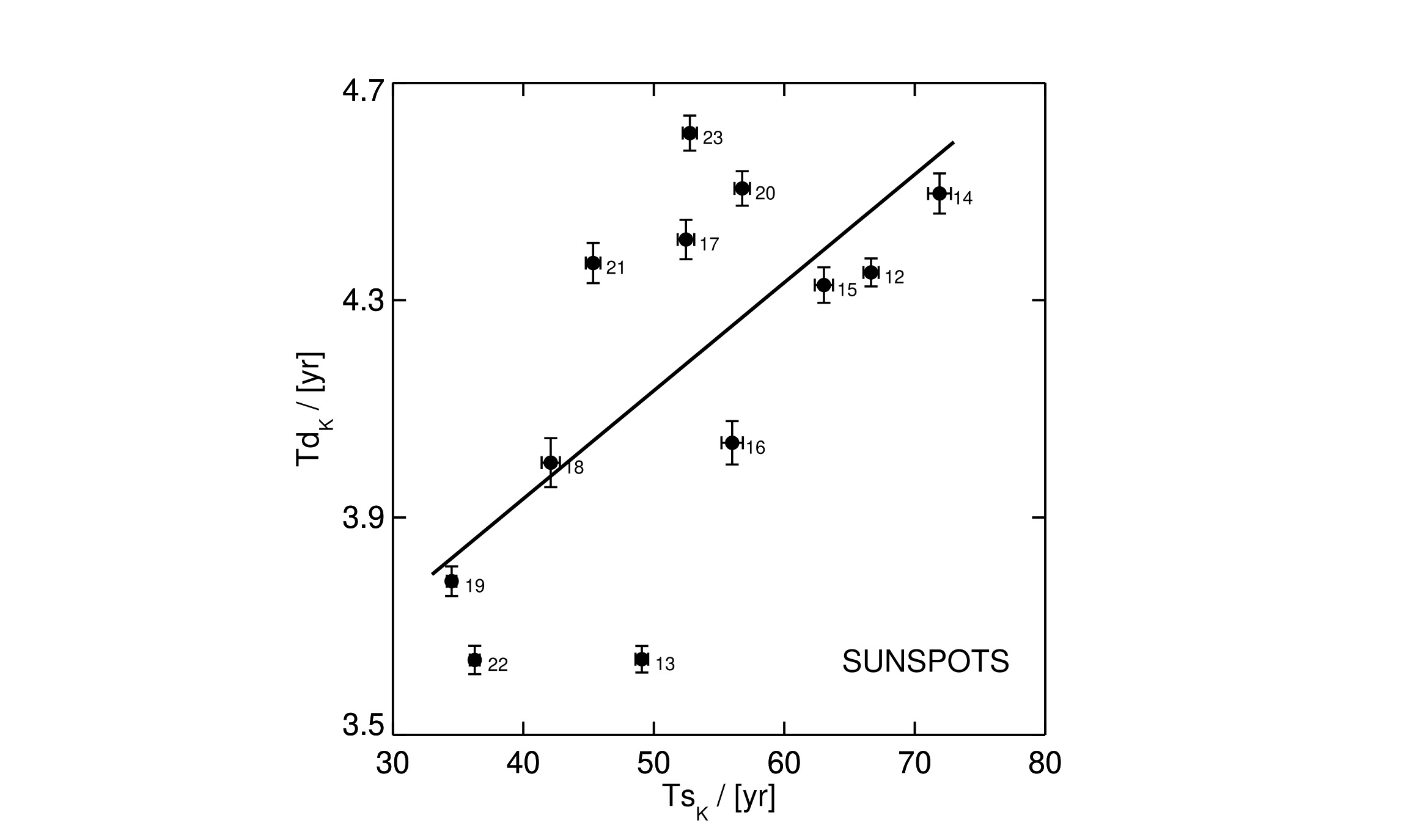}
\caption{
Relationship between the parameters $Td_k$ and $Ts_k$ for plage (left) and sunspot (right) areas.
The sunspot data include the additional cycles 12 and 13.
The filled circles, with 1-$\sigma$ error bars, show the data averaged over the individual 
solar cycle numbers. The regression lines, computed using a Bayesian method \citep{2007ApJ...665.1489K}, are given by Eq. \ref{TsvsTd}.}
\label{TsvsTd_plot}
\end{figure}

\begin{figure}[h]
\centering
\includegraphics[height=11 cm ,width=17 cm]{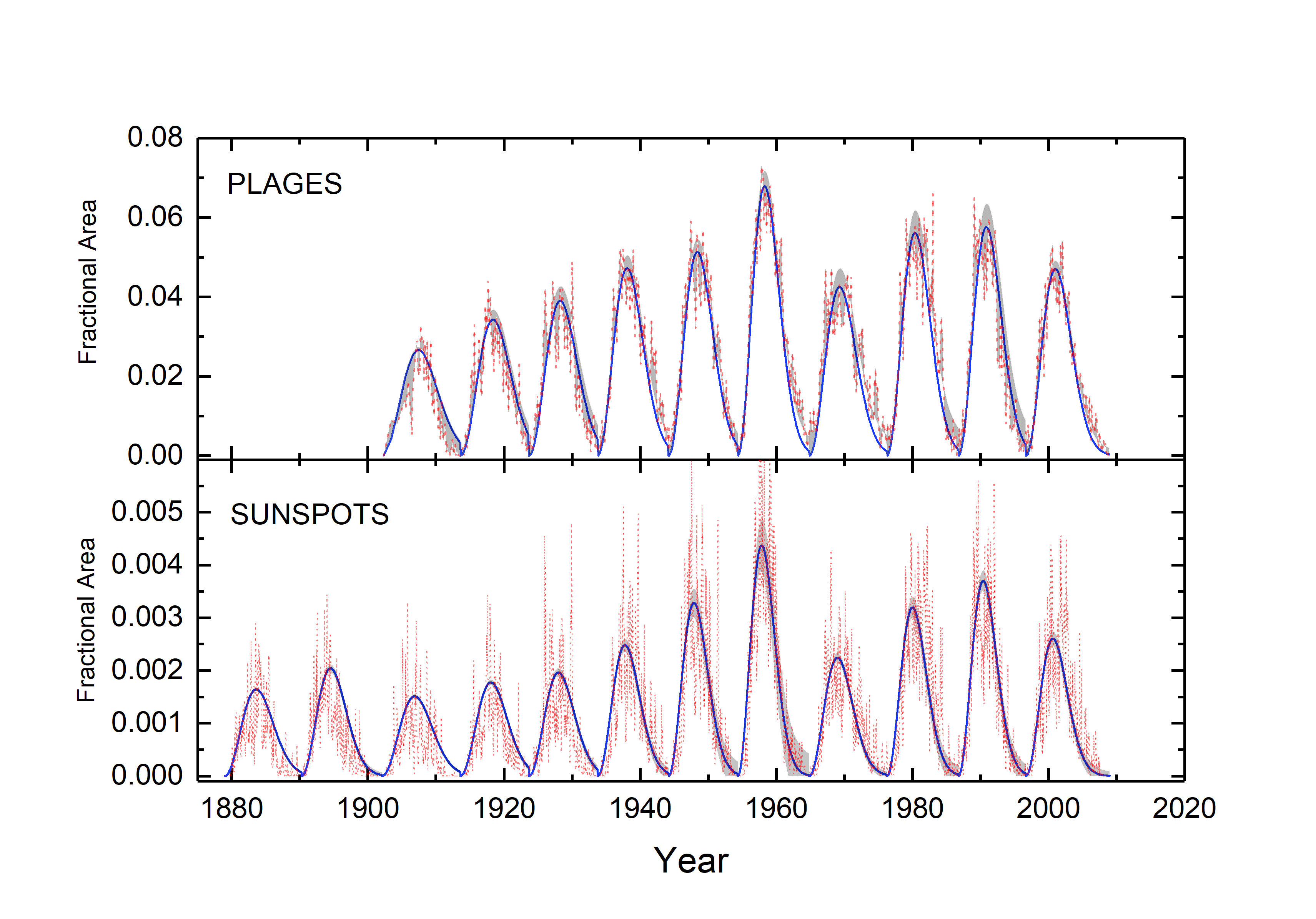}
\caption{\textit{Upper panel:} Parametric reconstruction of the plage coverage (blue line) with confidence range (gray). The plage composite from the MPI is shown for comparison (red line). \textit{Lower panel:} Parametric reconstruction of the Sunspot coverage (blue line) with confidence range (gray). \textbf{The reduction in the confidence range before 1940 is due to the lower SSN which makes it no longer appreciable.} The Sunspot composite from MPI is shown for comparison (red line).}
\label{AR_parametric}
\end{figure}

\begin{figure}[htbp] 
\centering 
\includegraphics[trim=4cm 0cm 4cm 0cm,clip]{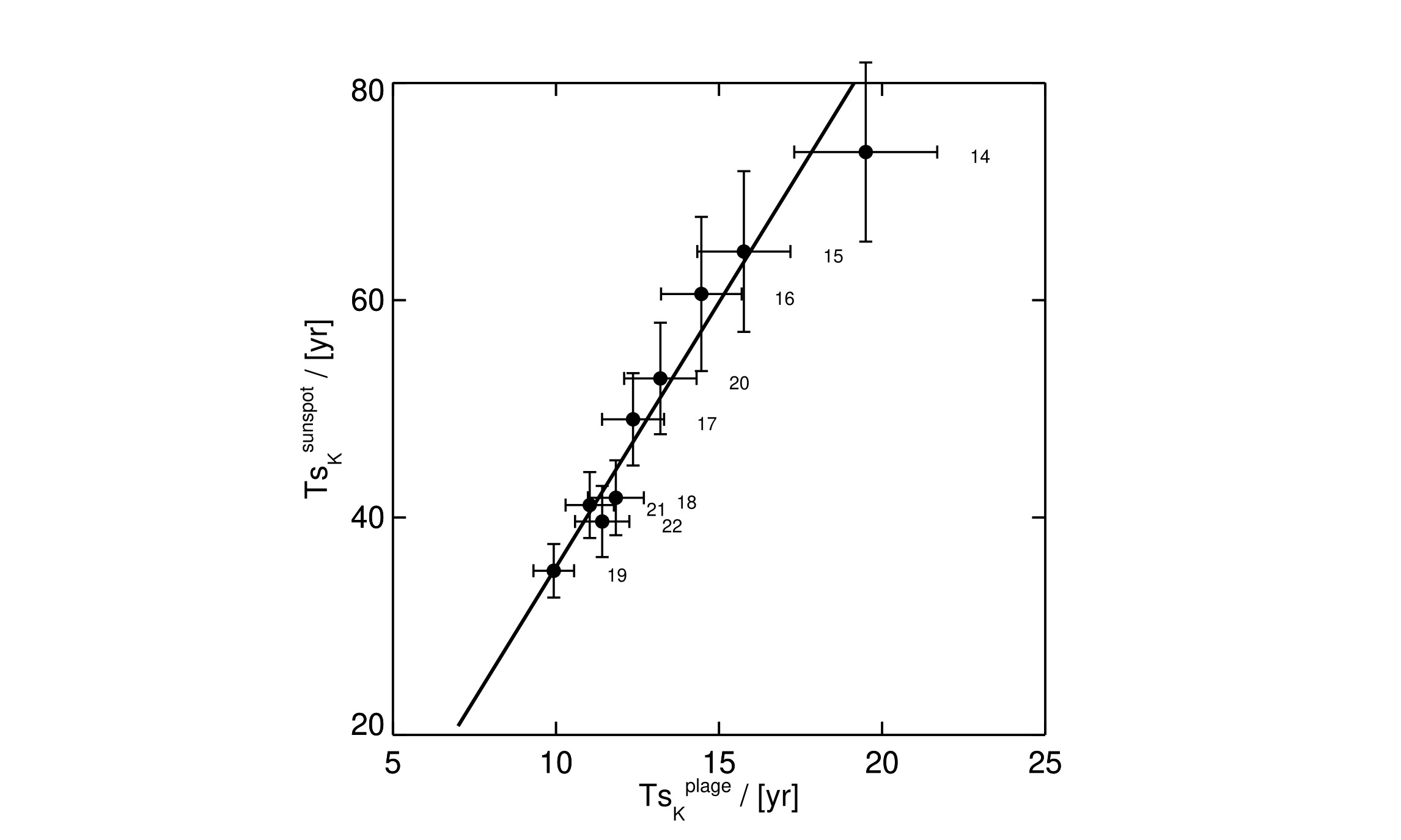} 
\caption{Correlation between $Ts_k^{sunspot}$, sunspot parameter, and $Ts_k^{plage}$, plage parameter. The error bars are 1-$\sigma$ standard error of the mean for both the sunspot and plage data. The regression line, computed using a Bayesian method (Kelly 2007), is given by Eq.\ref{Ts_spot_vs_Ts_plage}}. \label{Ts_spot_vs_phi} 
\end{figure}

\begin{figure}[htbp] 
\centering 
\includegraphics[trim=4cm 0cm 4cm 0cm,clip]{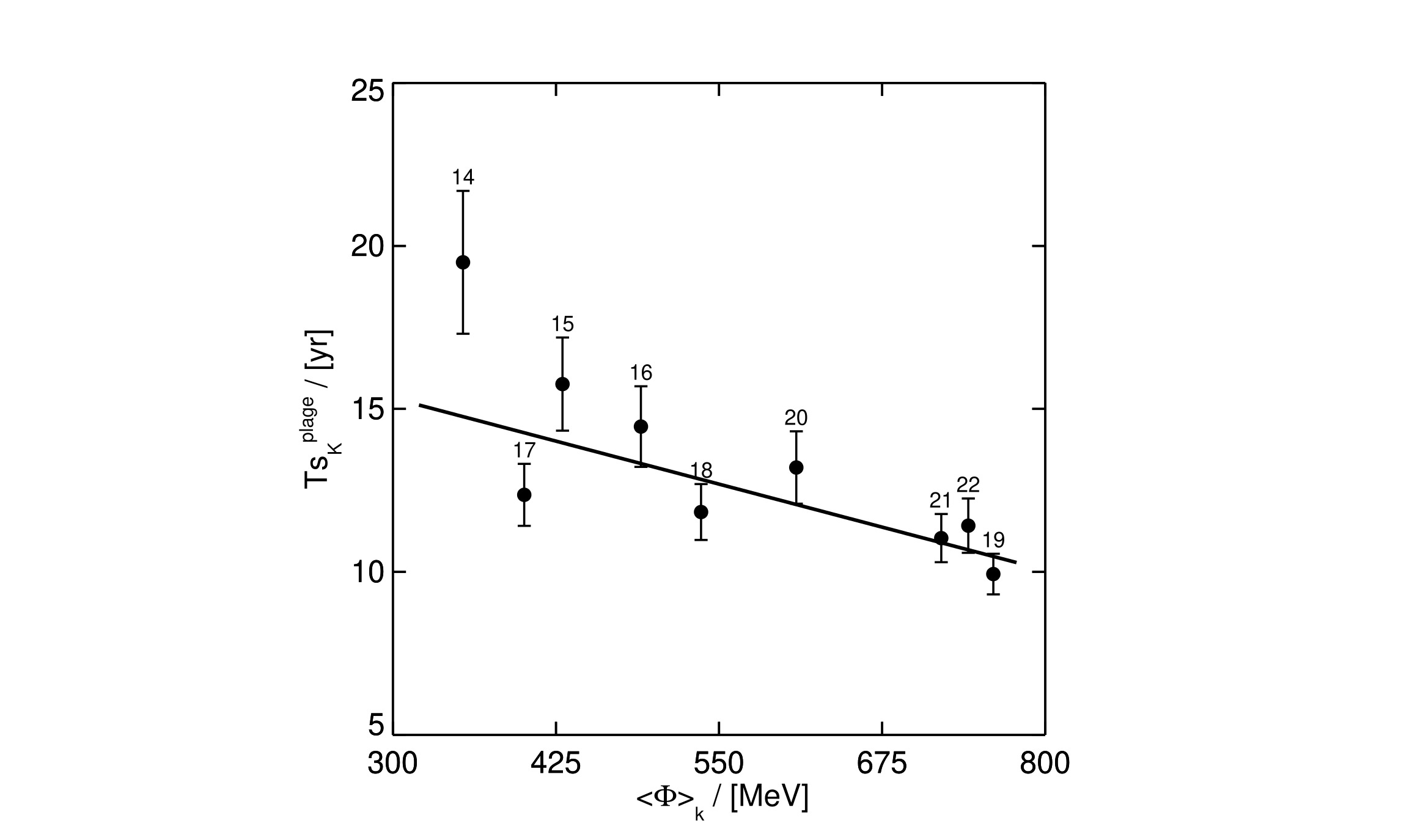} \qquad\qquad \caption{ Dependency of the parameter $Ts_k$ on the $k$-cycle averaged solar modulation potential $_k$. Individual solar cycle numbers are indicated in the plot, while the error bars are 1-$\sigma$ standard error of the mean. This relationship is well described by the linear model (solid line) given by Eq. \ref{Ts_plage_vs_phi_fit}. The Pearson's correlation coefficient is -0.77, at a confidence level greater than 95\%. } 
\label{Ts_plage_vs_phi} 
\end{figure}

As mentioned in the previous section, the first step of the proposed methodology consists in the estimate of plage and sunspot areas coverage at times when measurements were not available. Following the approach described in \cite{penza2021}, we characterize each cycle through the functional form given in \citet{Volobuev}
\begin{equation}
\label{cycle_form}
x_{k}(t) =  \left(\frac{t - T0_{k}}{Ts_{k}}\right)^{2} e^{-\left(\frac{t - T0_{k}}{Td_{k}}\right)^{2}} \quad \quad \textrm{for} \quad T0_{k} < t < T0_{k} + \tau_{k}
\end{equation}
where $T0_{k}$ is the time that marks the start of cycle $k$, and $Ts_{k}$ and $Td_{k}$ are two free parameters that describe the rising phase and the amplitude of the cycle. For this study we employed the starting times published in \citep{hathaway94, hathaway2015}.
In practice, because cycles with large amplitude are typically characterized by a shorter rising phase \citep[this is known as the Waldemeier effect,][]{hathaway94,hazra2015} $Ts_{k}$ and $Td_{k}$  are not independent variables. \citet{Volobuev} found that these two variables are linearly correlated.
We use the functional form in Eq.~\ref{cycle_form} to fit the sunspot and plage coverage data for all available solar cycles. The scatter plots of the derived $Td_{k}$ versus $Ts_{k}$ values are shown in Fig.~\ref{TsvsTd_plot} for both plages and sunspots. The error bars represent the uncertainties returned from the fitting procedure \citep{bevington}. 
Following \citet{Volobuev}, we perform a linear fit to the data.
Many methods have been proposed for performing linear regression when intrinsic scatter is present
and both variables are measured with error. Here we used the Bayesian approach 
described in \citet{2007ApJ...665.1489K} to determine the regression coefficients and 
their corresponding uncertainties.
We find that the dependence of $Td_{k}$ from $Ts_{k}$ is described by the following relationships: 
\begin{eqnarray}
\label{TsvsTd}
\nonumber Td_{k}^{plage} = (0.10 \pm 0.02) Ts_{k}^{plage} + (3.14 \pm 0.33)~~~ yr  \\
Td_{k}^{spot} = (0.02 \pm 0.01)Ts_{k}^{spot} + (3.14 \pm 0.43) ~~~  yr .      
\end{eqnarray}
The Pearson correlation coefficients between $Ts_{k}$ and $Td_{k}$ are r=0.81 and r=0.70 for plage and sunspot, respectively. To estimate the statistical significance of these correlations we performed a t-test and found there is a non-zero correlation between $Ts_{k}$ and $Td_{k}$, at a confidence level greater than 99\% for  plages and greater than 95\% for spots.
By inserting these relationships in Eq.~\ref{cycle_form}, we obtain a one-parameter functional form for the shape of the cycles:\\
\begin{eqnarray}
\label{cycle_form2}
\nonumber x_{k}^{plage}(t) =  \left(\frac{t - T0_{k}}{Ts_{k}}\right)^{2} e^{-\left(\frac{t - T0_{k}}{0.1 Ts_{k} + 3.14}\right)^{2}} \quad \quad \textrm{for} \quad T0_{k} < t < T0_{k} + \tau_{k} \\
x_{k}^{spot}(t) =  \left(\frac{t - T0_{k}}{Ts_{k}}\right)^{2} e^{-\left(\frac{t - T0_{k}}{0.02 Ts_{k} + 3.14}\right)^{2}} \quad \quad \textrm{for} \quad T0_{k} < t < T0_{k} + \tau_{k}
\end{eqnarray}
We use the new functional form in Eq.~\ref{cycle_form2} to repeat the fit procedure and to obtain two new dataset of $Ts_{k}$ values for sunspots and plages.
The temporal variation of the area coverages of plage and sunspot is therefore reproduced as summatory on cycle number k of the functional forms $x_{k}$. The parametric reconstructions for plage and sunspot area are reported in Fig.~\ref{AR_parametric}. The shaded areas in the figure represent the 1-$\sigma$ confidence level values estimated by taking into account the errors on the fit parameters.
This region is more evident in the plage case, while it is practically indistinguishable in the sunspots case. We note that both the reconstructed area coverages well reproduce the observed values.
It should also be noted that, as expected, $Ts_{k}^{plage}$ and $Ts_{k}^{spot}$ are strongly correlated. The Pearson correlation coefficient is 0.98 and, as shown in Fig.~\ref{Ts_spot_vs_phi}, their dependence is well described by the following linear relationship:
\begin{equation}
\label{Ts_spot_vs_Ts_plage}
\nonumber Ts_{k}^{spot} = (4.9 \pm 1.4) Ts_{k}^{plage} - (13.3 \pm 16.4) ~~~yr
\end{equation}
The parameterization described above allows us to characterize each cycle through a single parameter ($Ts_{k}$). We then use the solar modulation potential to reconstruct the shape of each solar cycle for periods when plage and sunspot measurements were not available.
In Fig.~\ref{Ts_plage_vs_phi} we show the scatter plot between the plage parameter $Ts_{k}^{plage}$ and the solar modulation potential $<\Phi>_{k}$  for each cycle $k$. Here $<\Phi>_{k}$ is the averaged-integral value of the solar modulation potential over each cycle. The dependency of the parameter $Ts_{k}^{plage}$ on the solar modulation potential $<\Phi>_{k}$ is well described by the following relation:
\begin{equation}
\label{Ts_plage_vs_phi_fit}
\nonumber Ts_{k}^{plage} = -(0.011 \pm 0.002)<\Phi>_{k} + (18.5 \pm 1.5)~~~ yr
\end{equation}
The Pearson's correlation coefficient is -0.77, at a confidence level greater than 95\%.
\begin{figure}[htbp]
\centering
\includegraphics[width=0.9\linewidth]{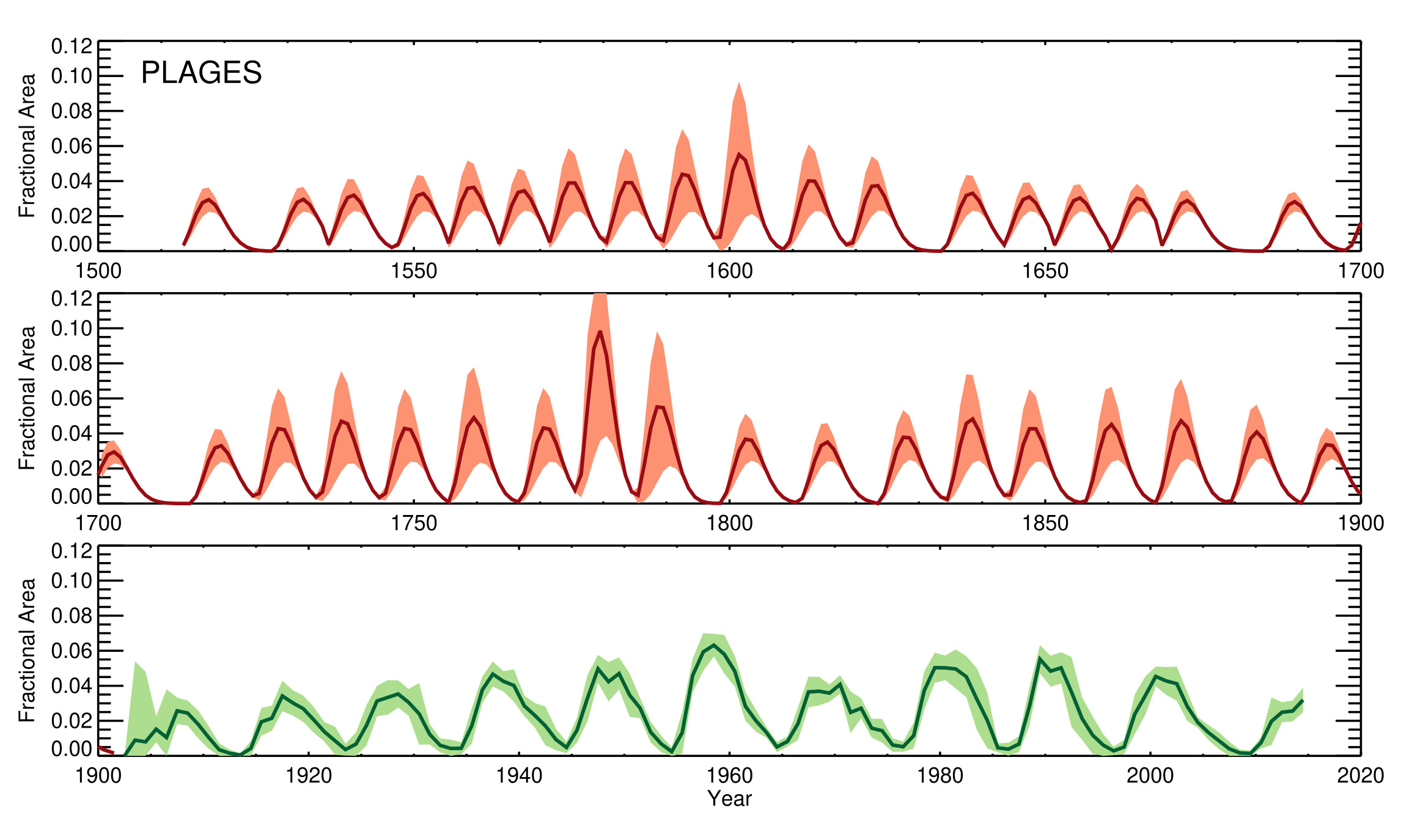}
\caption{Annual values of the fractional solar disk coverage by plage as a function of time.
This composite is derived by merging the values from a reconstructed time series using Eq. \ref{Ts_plage_vs_phi_fit} (red solid line) and observed data (green solid line). The light-red region indicates the uncertainty range in the reconstructed time series, while the light-green region shows the 5$\sigma$ confidence interval of the mean annual value.
\label{fac_rec}}
\end{figure}
We use Eq.~\ref{Ts_plage_vs_phi_fit} to reconstruct the plage coverage trend back in time. We decided to consider the $\Phi$ data starting from 1513, because before this date the sampling of the solar modulation potential data no longer shows the 11-year cycle modulation
due to a much lower temporal sampling from the IntCal04 calibration curve as reported in \cite{muscheler2007}.\\ 
In Fig.~\ref{fac_rec} we plot the composite of the plage coverage, that consists of the reconstructed plage area obtained using Eq.~\ref{Ts_plage_vs_phi_fit} (for the period 1513-1902) and the original observed data (from 1902 to 1996). The reconstruction is shown as a "strip" that takes into account the error propagation of the Ts values in the functional form in Eq.~\ref{cycle_form}.
The values of cycle start data $T0_{k}$ and of cycle duration $\tau_k$, that are not free parameters, are imposed from 1755, i.e. from Cycle 1, by using the data available at SILSO website \footnote{The table of minima, maxima and cycle durations are available on SILSO link: https://wwwbis.sidc.be/silso/cyclesmm.}, and \cite{hathaway2015} while before 1755 by using directly the trend of the solar modulation potential. The values are reported in Table~\ref{tab:1}.

\begin{deluxetable*}{llllll}[h!]
\label{tab:1}
\tablenum{1}
\tablecaption{Solar Cycle list \label{tab:SC}}
\tablewidth{0pt}
\tablehead{
\colhead{Solar Cycle} & \colhead{T0 (yr)} & \colhead{$\tau$ (yr)} & \colhead{Solar cycle} & \colhead{T0 (yr)} & \colhead{$\tau$ (yr)} \\
}
\startdata
 -24 &  1502.5 & 10.0  &   -  &     -     & -       \\
 -23 &  1512.5 & 15.0  &  1 &  1755.17  & 11.33 \\
 -22 & 1527.5  & 8.0  &  2 &  1766.50 & 9.00 \\
-21 &  1535.5  & 11.0  &  3 &  1775.50 & 9.25\\
-20 &  1546.5  & 8.0 &  4 &  1784.75 & 13.58 \\
-19 &  1554.5  & 8.0  &  5 &  1798.33 & 12.33 \\
-18 &  1562.5  & 8.0  &  6 &  1810.67 & 12.75\\
-17 &  1570.5  & 8.0  &  7 &  1823.42 & 10.50\\
-16 &  1579.5  & 9.0  &  8 &  1833.92 & 9.67 \\
-15 &  1588.5  & 9.0  &  9 &  1843.583 & 12.40\\
-14 &  1597.5  & 9.0  & 10 &  1855.98 & 11.27\\
-13 &  1608.5  & 11.0  &  11 &  1867.25 & 11.73\\
-12 &  1618.5  & 10.0  & 12 &  1878.98 & 11.27\\
-11 &  1633.5  & 15.0  &  13 &  1890.25 & 11.83\\
 -10 &  1642.5  & 11.0  &   14 &  1902.08 & 11.50\\
-9 &  1650.5  & 8.0  &  15 &  1913.58 & 10.08\\
-8 &  1660.5  & 10.0  &  16 &  1923.67 & 10.08\\
 -7 &  1667.5  & 7.0 &  17 &  1933.75 & 10.42 \\
 -6 &  1684.5  &  17.0 & 18 &  1944.17 & 10.17\\
 -5 &  1697.5  & 13.0  &  19 &  1954.33 & 10.50\\
 -4 &  1714.5  & 17.0  &  20 &  1964.17 & 11.42\\
 -3 &  1724.5  & 10.0  &  21 &  1976.25 & 10.50\\
 -2 &  1734.5  & 10.0  &  22 &  1986.75 & 9.92 \\
 -1 & 1744.5  &  10.67  &  23 &  1996.67 & 12.30 \\
\enddata
\tablecomments{In this table are reported the start data (T0) and the duration ($\tau$) of the Solar Cycles}
\end{deluxetable*}

\begin{figure}[htbp]
\centering
\includegraphics[width=0.9\linewidth]{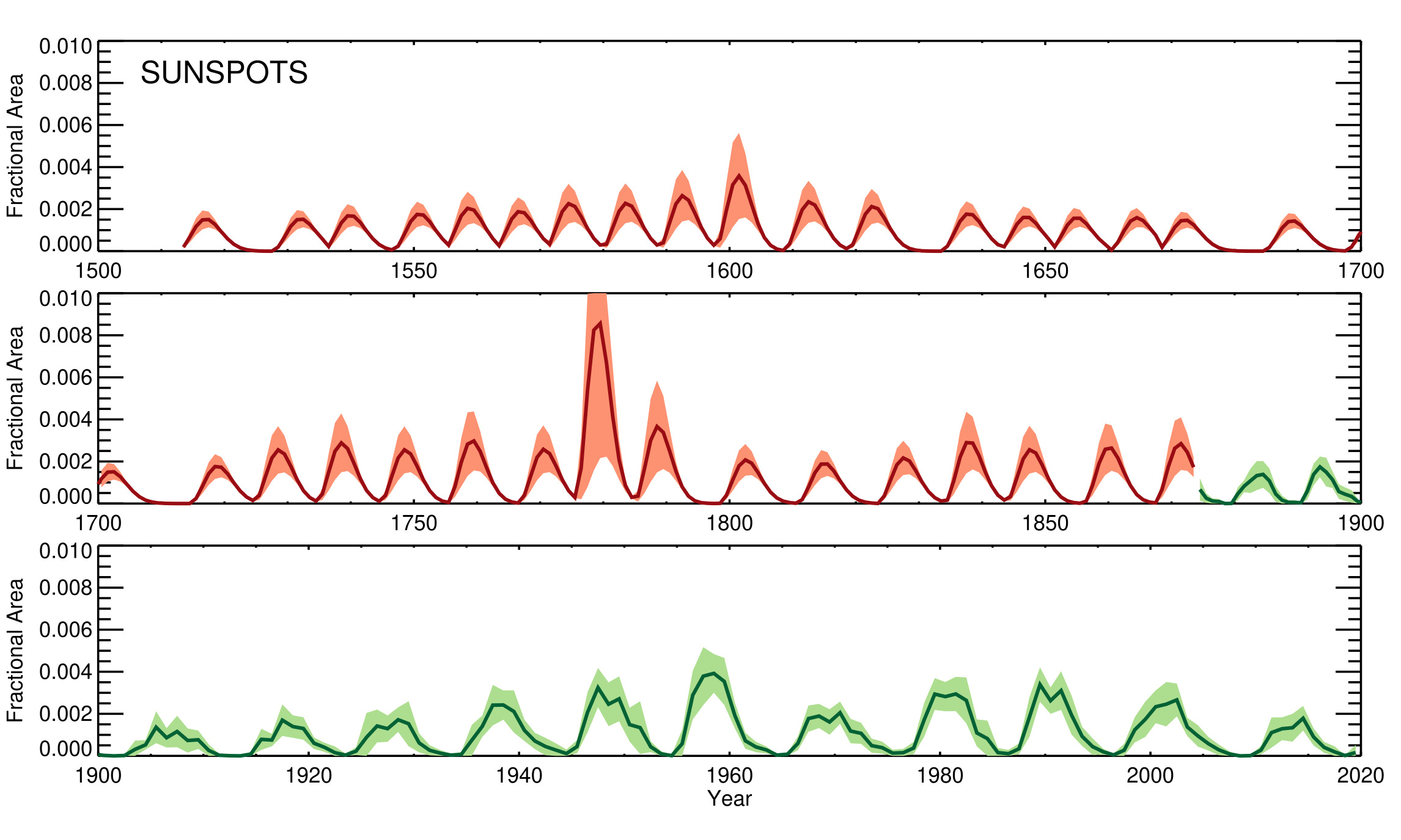}
\caption{
Same as in Fig. \ref{fac_rec}, but for sunspot data. Here, the light-green region shows the 20$\sigma$ confidence interval
of the mean annual value.}
\label{spot_rec}
\end{figure}

In order to reconstruct the sunspot area coverage back to 1513 we used the linear relation between $Ts_{k}^{plage}$ and $Ts_{k}^{spot}$ described in Eq.~\ref{Ts_spot_vs_Ts_plage}.   
The resulting composite of sunspot area coverage is shown in Fig.~\ref{spot_rec}.
\section{Decomposition of solar modulation potential $\Phi$} 
\label{sec:sec4}      
\begin{figure}[!htb]
\centering
\includegraphics[height=9 cm ,width=19 cm]{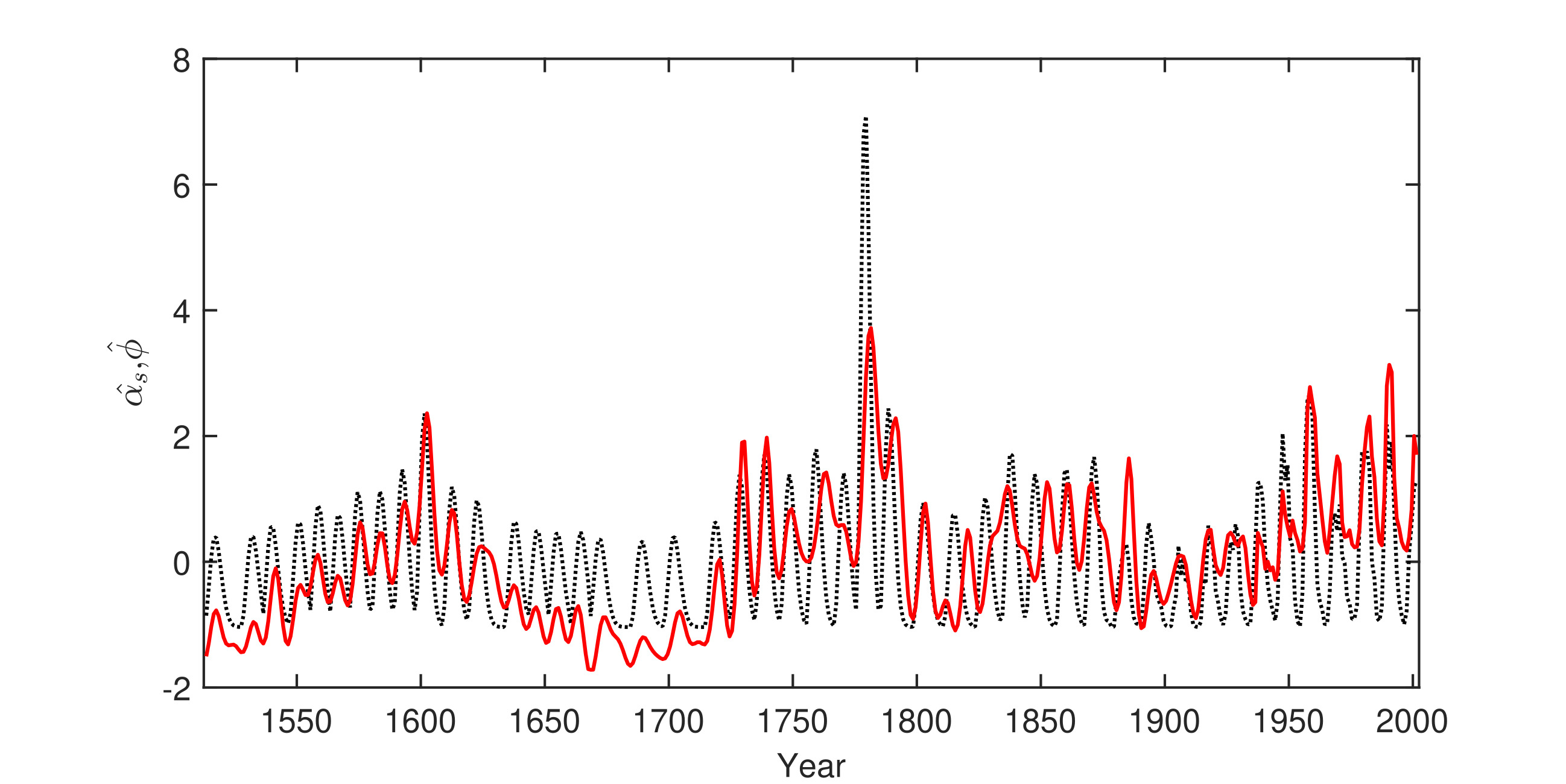}
\caption{Annual values of the standardized fractional solar disk coverage by sunspots $\hat{\alpha}_{s}$ (dotted black line) and standardized potential $\hat{\Phi}$ (continuous red line). The sunspots signal has been reconstructed until the 1873, while more recent values are recorded \citep{mandal2020}.}
\label{potential_vs_spot}
\end{figure}

In the introduction of this paper we discussed the joint contribution of the quiet Sun and of photospheric magnetic structures, i.e., sunspots, faculae, network, to the TSI variability. Although we know that the contribution of the quiet Sun's magnetic field is negligible, at least on the scale of ten years of the solar cycle, we cannot exclude that variations of the quiet Sun could play a role in modulating the TSI on longer temporal scales. To estimate the long-term modulation in the TSI and separate the possible contributions to the TSI of the different solar magnetic structures we resort to a decomposition of the temporal modes present in the solar modulation potential, which, as already stated in Sec.~\ref{sec:sec1}, we assume to be a proxy of solar surface magnetism.
In order to separate the different time-scale components of the solar modulation potential $\Phi$ we employ the Empirical Mode Decomposition (EMD) algorithm \citep{1998RSPSA.454..903H}.
The EMD algorithm is an adaptive and efficient signal decomposition technique, based on the local characteristic time scale of the signal, suitable for nonlinear, chaotic or non-stationary processes.
The EMD's intent is to decompose and simplify the original signal into a set of so-called intrinsic mode functions (IMFs) including a monotonic residual. Each IMF, representative of a mode embedded in the data, is calculated \textit{a posteriori} from the signal by a procedure called the sifting process \citep[for more details refer to][]{1998RSPSA.454..903H}.\\
The first step in applying the EMD algorithm to our problem consists in the standardization of the solar potential $\Phi$ and of the fractional solar disk coverage by sunspots $\alpha_s$. 
The standard score, or Z-score, is calculated as:
 \begin{equation}
  z=\frac{x-\mu}{\sigma}   
  \label{standardization}
 \end{equation}
 where z is the dimensionless standardized signal, x is the original signal (i.e., $\phi$ and $\alpha_{s}$), 
 $\mu$ and $\sigma$ the mean and the standard deviation, respectively. 
 Shown in Fig.~\ref{potential_vs_spot} are the standardized potentials $\hat{\phi}$ and the standardized sunspots coverage area $\hat{\alpha}_{s}$.
To identify and extract the long-term components present in the $\hat{\Phi}$ signal we use the EMD procedure as described in the
next section.
\subsection{Estimate of $\hat{\Phi}$ long-term component based on EMD}
\begin{figure}[h!]
\centering
\includegraphics[height=11 cm ,width=18 cm]{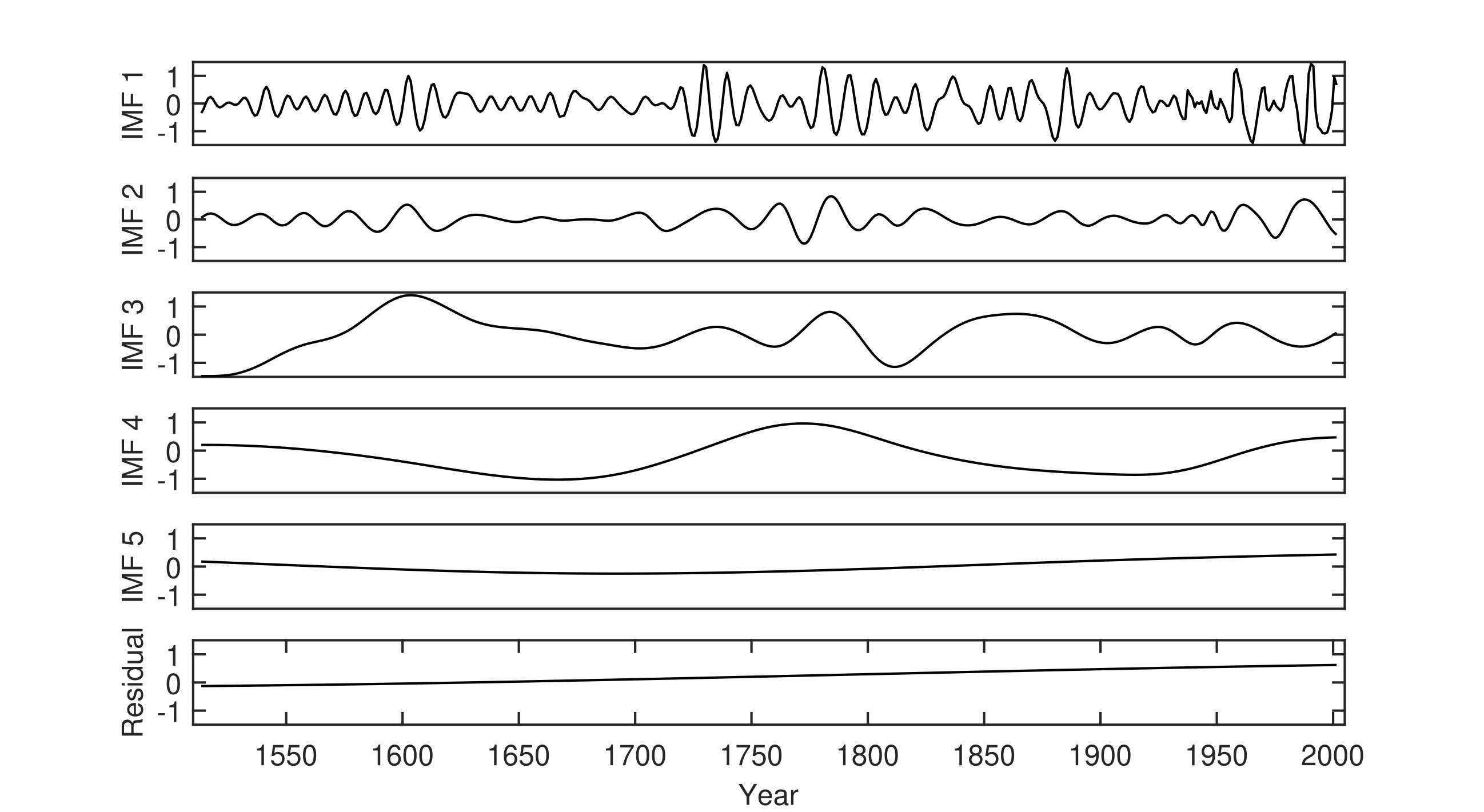}
\caption{Empirical Mode Decomposition of the standardized potential $\hat{\Phi}$ . The IMFs and the residual are shown.IMF1 and IMF2 capture the shortest periods, i.e., periods of approximately 11 years and 22 years, respectively. The residual shows the monotonic trend of the signal. The ordinates have the same scale, so the IMFs can be easily compared. }
    \label{fig:EMD_decomposition}
\end{figure}

Recall that the goal of EMD algorithm is to decompose a signal into a finite number of basis functions, the IMFs, and a residual $R_{n}(t)$. A generic signal X(t) is therefore decomposed as:
\begin{equation}
    X(t)=\sum_{i}^{n-1} IMF_{i}(t)+R_{n}(t)
\end{equation}
IMFs are oscillatory modes (or components) whose amplitude and period can vary over time, while the residual R(t) is the monotonic trend of the signal. Unlike other signal analysis methods, such as Fourier analysis or wavelets analysis, the basis functions in EMD are not predefined but are empirically calculated from the signal by the algorithm. 
For these characteristics the EMD method is now widely used in physics and engineering. 
In solar physics EMD has often been used for its ability to decompose signals that are not simply periodic \citep[e.g.][]{Barnhart2011,2014A&A...569A.102S,Kolotkov2016,lovric2017,Keys2018A,Vecchio2019,2020AnGeo..38..789B,Lee2020}.\\
The IMFs of the standardized potential $\hat{\Phi}$ are shown in Fig.~\ref{fig:EMD_decomposition}.
The first two components, IMF1 and IMF2, show periodicity associated with the 11-year and 22-year solar magnetic cycles,
while the remaining IMFs show longer periods. The residual function shows the monotonic trend of the signal. 
It is worth remembering
that, differently from the sine and cosine components of the Fourier transform, the IMFs represent oscillatory modes whose amplitude and period can vary over time.\\
At this point it is possible to estimate the secular variation $\Phi_{LT}$ by eliminating the common contributions described by the low-order IMFs of $\hat{\Phi}$ and $\hat{\alpha}_{s}$.
Since we want to extract only the long-term solar modulation  $\Phi_{LT}(t)$ from $\hat{\Phi}$, we exclude IMF1 and IMF2 from the reconstruction. Therefore, $\Phi_{LT}(t)$ is obtained from
\begin{equation}
    \Phi_{LT}(t)=\sum_{i=3}^{5}C_i \cdot IMF_i(t)+C_R \cdot R(t),
    \label{modulation}
\end{equation}
where $C_i$ and $C_R$ are parameters which establish the weight of the different normalized IMFs, including the residual $R(t)$, in the final modulation.
These parameters are defined so that $\Phi_{LT}(t)$ is the modulation present in $\hat{\Phi}$ but not in $\alpha_{s}$.
Therefore, we minimize the residuals between $\hat{\Phi}(t)$ and $\hat{\alpha}_{s}(t)+\Phi_{LT}(t)$ to estimate the free parameters $C_i$ and $C_R$ 
so that $\Phi_{LT}$ represents the secular modulation of the solar modulation potential without the contribution of the modes common with $\hat{\alpha_{s}}$.\\
To get the best estimate of the parameters we use a Bayesian approach. More in detail, 
we use a Monte Carlo Markov Chain (MCMC) algorithm for sampling probability density functions. 
The Bayes theorem states that: 
\begin{equation}
    p(\mathbf{C}|\hat{\Phi}(t))=\frac{p(\hat{\Phi}(t)|\mathbf{C}) \cdot p(\mathbf{C})}{p(\hat{\Phi}(t))} 
\end{equation}
where: \textit{i)} $p(\mathbf{\mathbf{C}}|\hat{\Phi}(t))$ is the posterior probability, i.e., the probability of the model of having true parameters $\mathbf{\mathbf{C}}$ given the observed data $\hat{\Phi}(t)$, \textit{ii)} $p(\hat{\Phi}(t)|\mathbf{C})$ is the likelihood, i.e., the probability of having $\hat{\Phi}(t)$ given the set of parameters $\mathbf{C}$, and \textit{iii)} $p(\mathbf{C})$ is the prior probability, i.e., the bias on the possible values of the parameters coming from a-priori knowledge. In our analysis, $\mathbf{C}$ are the parameters that we want to estimate ($C_{3},C_{4},C_{5},C_{R}$) and $\hat{\Phi}(t)$ the vector of data (i.e., the standardized potential). \\ 
Since MCMC algorithms do not depend on the evidence $p(\hat{\Phi}(t))$, by imposing a flat prior probability and a Gaussian likelihood function we have
\begin{equation}
   \text{ln} \,  p(\edit1{\theta|\hat{\Phi}})= -\frac{1}{2} \sum_{t=1}^{N} (\hat{\Phi}(t) -\Phi_{LT}(t)-\hat{\alpha}_{s}(t))^{2}
\end{equation}
where t is the discrete time, in years, over which the functions are estimated,
and N=489 is equal to the total number of years used (discrete time-domain).
The MCMC algorithm used here is the Goodman-Weare \citep{2010CAMCS...5...65G}, implemented in Python \citep{2013PASP..125..306F}.
To estimate the most probable values of the different weights $C_i$ and estimate their confidence interval at $1\sigma$ we performed a suitable number of tests (i.e., 5000 iterations and 50 chains). The calculated values for the $C_i$ are:
$C_{3}=0.33\pm0.05$, $C_{4}=0.26\pm0.05$, $C_{5}=0.22\pm0.05$, and $C_{R}=0.15\pm{0.04}$, respectively.
\section{TSI reconstruction} 
\label{sec:sec5} 
Like other methods presented in the literature (see Sec.~\ref{sec:intro}), our TSI reconstruction is based on the assumption that irradiance is modulated by solar surface magnetism. Specifically, we assume that the solar irradiance F at time $t$ is given by:
\begin{equation}
 \label{F_rec}
F(t) =  \sum_{j} \alpha_{j}(t) F_{j},
\end{equation}
where $F_{j}$ is the contribution to the TSI from the  j-feature (quiet, network, facula, and sunspot) and $\alpha(t)_{j}$ is the respective coverage. 
We use the reconstructed plage coverage area as proxy of the facular coverage $\alpha_{f}(t)$ (see \cite{2010RvGeo..48.4001G}). We further assume that only the surface coverages change with the time, while the average contrasts are time-independent.
We rewrite the Eq.~\ref{F_rec} by expliciting the network, facular and sunspot contributions:
\begin{equation}
 \label{F_rec2}
 F(t) =  F_{q} (1- \alpha_{n}(t) - \alpha_{f}(t) - \alpha_{s}(t)) + \alpha_{n}(t) F_{n} + \alpha_{f}(t) F_{f} + \alpha_{s}(t) F_{s} 
\end{equation}
where the subscripts $n$, $f$ and $s$ indicate network, facular and sunspot component, respectively. 
We adopt a linear correlation between $\alpha_{n}$ and $\alpha_{f}$ \citep[e.g.][]{Criscuoli2018, pooja2021}:
\begin{equation}
 \label{an-af}
 \alpha_{n} = (A_{n} +  B_{n} \alpha_{f}(t)).
\end{equation}
So that
\begin{equation}
 \label{F_rec3}
 F(t)  = F_{q} + (F_{n} - F_{q})(A_{n} +  B_{n} \alpha_{f}(t)) + (F_{f} - F_{q}) \alpha_{f}(t) + (F_{s} - F_{q}) \alpha_{s}(t).
\end{equation}
The equation can be simplified if we rewrite it as a relative variation with respect to the radiative flux of the Quiet Sun:
\begin{equation}
\label{deltaF}
\Delta F(t) = \frac{F(t) - F_{q}}{F_{q}} = A_{n} \frac{F_{n} - F_{q}}{F_{q}} + \alpha_{f}(t) B_{n} \frac{F_{n} - F_{q}}{F_{q}} + \alpha_{f}(t) \frac{F_{f} - F_{q}}{F_{q}} + \alpha_{s}(t) \frac{F_{s} - F_{q}}{F_{q}} 
\end{equation}
\begin{equation}
\label{deltaF}
\Delta F(t) = A_{n} \delta_{n}+ \alpha_{f}(t)(B_{n} \delta_{n} + \delta_{f}) + \alpha_{s}(t) \delta_{s} = C_{n} + \alpha_{f}(t) \delta_{fn} + \alpha_{s}(t) \delta{_s} 
\end{equation}
where $C_{n}$ is a constant and represent the product between the network contrast and the network coverage when the facular coverage is zero, 
$\delta_{fn}$ is a linear combination of network and facular relative contrast, while $\delta_{s}$ is the sunspot relative contrast. The values of $C_{n}$, $\delta_{fn}$ and $\delta_{s}$ are estimate by best fitting Eq.~\ref{deltaF} with measurements of TSI variability \footnote{specifically we use the composite available at https://www.pmodwrc.ch/en/research-development/solar-physics/tsi-composite/} along Solar Cycle 22 and 23. We fit the two cycles separately in order to consider the long-term component negligible, and for each cycle we use the corresponding measured $\alpha_{s}$ and $\alpha_{f}$. As values for the TSI reconstruction we take the average of the two set of $C_{n}$, $\delta_{fn}$ and $\delta_{s}$ estimated from the two fits:\\
\begin{figure}[htbp]
\centering
\includegraphics[height=7cm, width=12 cm]{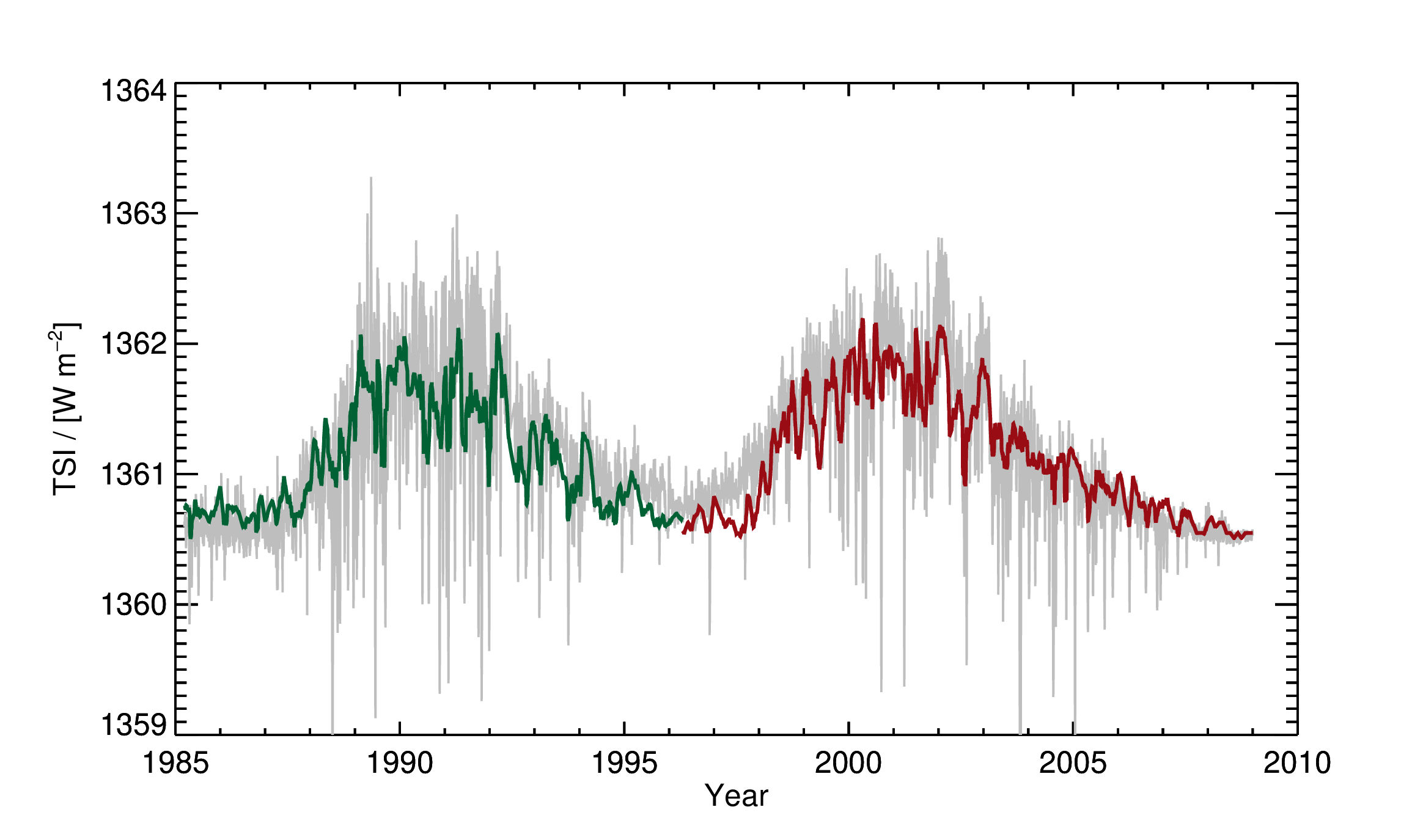}
\qquad\qquad
\caption{Comparison between the TSI PMOD-composite (in grey) and the reconstructed TSI time series for solar cycles 22 (green) and 23 (red). The TSI reconstruction was computed independently for cycle 22 and 23 (see text).}
\label{TSI_fit2}
\end{figure}
\begin{eqnarray}
\nonumber C_{n} = \frac{C_{n}^{(22)} + C_{n}^{(23)}}{2} = \frac{(1.28 ^{.} 10^{-3} \pm  5 ^{.} 10^{-5}) + (1.34 ^{.} 10^{-3} \pm  7 ^{.} 10^{-5})}{2} = 1.31 ^{.} 10^{-3} \pm  6 ^{.} 10^{-5} \\
\nonumber \delta_{fn} = \frac{\delta_{fn}^{(22)} + \delta_{fn}^{(23)}}{2} = \frac{(0.029 \pm  0.005) + (0.025 \pm  0.003)}{2} =  0.027 \pm 0.004 \\
\nonumber \delta_{s} = \frac{\delta_{s}^{(22)} + \delta_{s}^{(23)}}{2} = \frac{(-0.16 \pm 0.05) + (-0.18 \pm  0.07)}{2} = - 0.17 \pm 0.06 \\
\end{eqnarray}
The measured and reconstructed TSI for Cycle 22 and Cycle 23 are shown in Fig.~\ref{TSI_fit2}.
The TSI absolute values are obtained by setting the $F_q$ value to the TSI value measured at the 1986 minimum. 
The choice of this minimum as the baseline for normalization has no substantial effect in the final reconstruction, as the fluxes of measured minima differ no more than 0.2$Wm^{-2}$.\\ 
Although we have imposed no constraints on the model, we note that the values of the fitted parameters are not significantly different from others reported in the literature. For example, our result of $C_{n}$ is consistent with results of \cite{foukal1991}, that provide a value of the network contrast weighted by corresponding fractional area at the solar minimum of about $1.08 \, ^{.} 10^{-3} $. Similarly,  the value of $\delta_{fn}$ represents a sensible mixed facular and network bolometric contrast \citep[e.g.][]{foukal2004}. The sunspot bolometric contrast, however, results about 50\% less intense than the values reported in the literature \citep[e.g.][]{chapman1994,walton2003}.\\
In order to reproduce the TSI over temporal scales longer than the decadal one, we modulate the network component ($C_{n}$), present even in the absence of other magnetic structures. In this way, we attribute to this parameter all the effects of the open or "hidden" magnetic field and we rewrite the Eq. \ref{deltaF} as
\begin{equation}
\label{deltaF2}
\Delta F(t) = C_{n} mod_{\Phi}(t) + \alpha_{f}(t) \delta_{fn} + \alpha_{s}(t) \delta{_s}
\end{equation}
To derive the modulation function $mod(\Phi)$, we consider the residual composition of IMFs of $\Phi$ as explained in Section \ref{sec:sec5}.
The modulation function $mod(\Phi)$ is the long-term component of the solar modulation potential $\Phi_{LT}$ described in Sec.~\ref{sec:sec4}, properly normalized.
We impose a normalization parameter as following:
\begin{equation}
mod_{\Phi}(t) = (\Phi_{LT}(t)+1)N_{ref}  
\end{equation}
and determine the parameter $N_{ref}$ by best fit optimizing the comparison with the entire PMOD TSI composite of TSI and obtaining the value $N_{ref}  = 0.38 \pm  0.02$.\\
Eventually, we can reconstruct the behavior of the TSI in the period (1513-2001). The estimated TSI, reconstructed using the approach proposed in this work, is shown
in Fig.~\ref{TSI_rec}, together with the TSI proposed by \cite{wu2018}, SATIRE, \cite{egorova2018}, CHRONOS, and \cite{coddington2019}, NRL-TSIv2.
\begin{figure}[htbp]
\centering
\includegraphics[height=13.5 cm, width=19 cm]{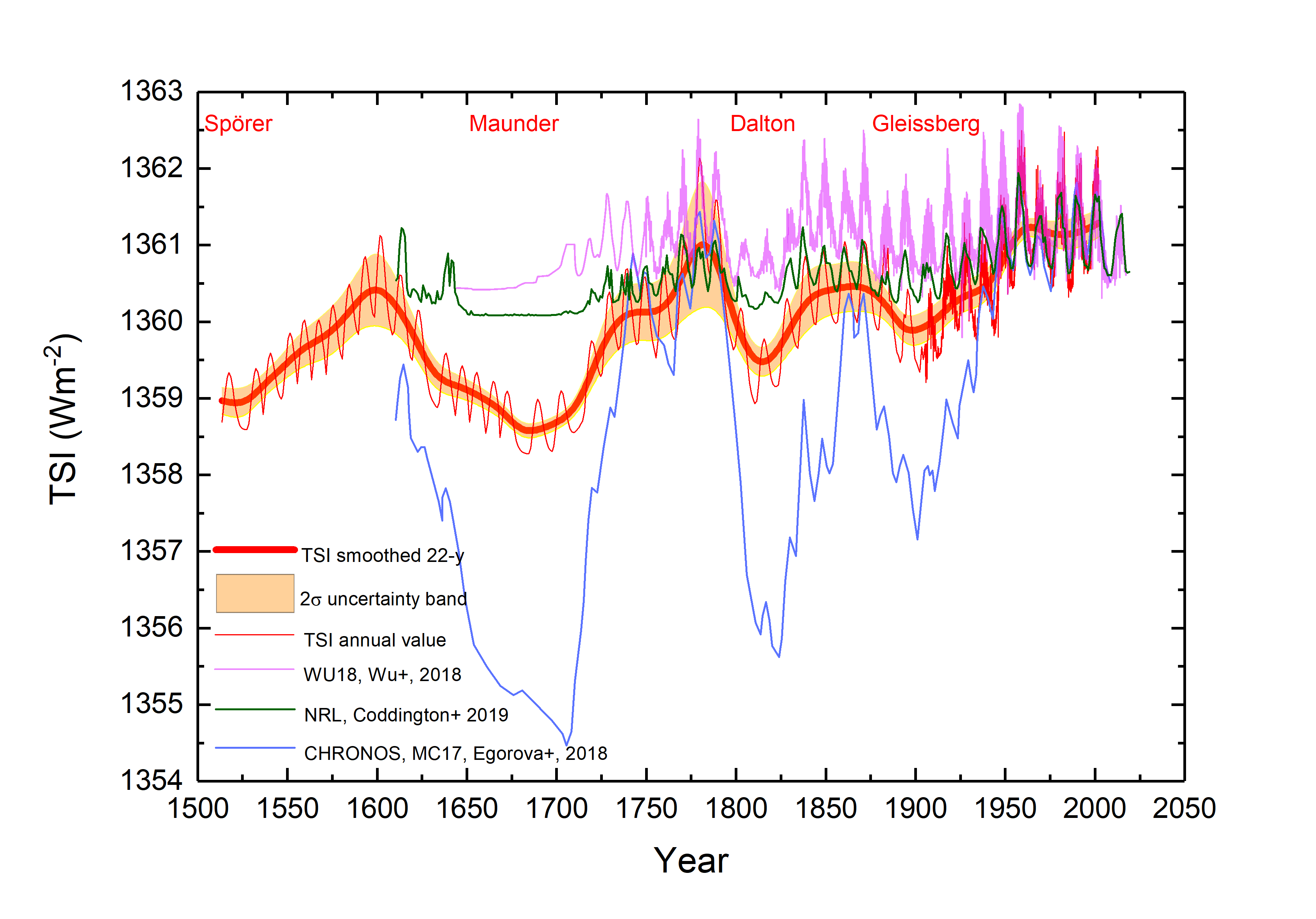}
\qquad\qquad
\caption{Four reconstructions of TSI are reported, obtained with different methods for comparison purposes only. Our TSI reconstruction is shown as a continuum red line; the red bold line is the TSI smoothed over a period of 22 yr with a $2-\sigma$ uncertainty orange stripe. The NRL TSI data Climate Data Record (green line) from \citet{coddington2019}. TSI values from \citet{wu2018} (magenta). CHRONOS, MC17 TSI values from  \citet{egorova2018} (blue). The names refer to the grand minima that occurred in the analyzed period, i.e., Spörer, Munder, Dalton and Gleissberg.}
\label{TSI_rec}
\end{figure}
\section{Discussion and Conclusions}\label{sec:sec6}

In this work we present a reconstruction of the total solar irradiance variability from the pre-industrial era to the present.
Our approach uses modern TSI measurements to estimate the contribution of sunspots and faculae in terms of area coverages and bolometric contrast, and the solar modulation potential to take into account for the long-term (longer than 22 years) variation of a quiet component. 
It's worth noting that our estimates of area coverage are in agreement with the values derived directly from the observations of sunspots and faculae (see Fig.~\ref{fac_rec} and Fig.~\ref{spot_rec}). Moreover, our estimated values for the bolometric contrast are in agreement with values reported in literature (for a complete discussion see Sec. \ref{sec:sec5}).\\
Regarding TSI, our reconstruction produces a variation from the Maunder minimum to the present of approximately 2.5 $Wm^{-2}$. This value is in between those obtained with models that do not take into account secular variations of a quiet component, such as SATIRE \citep{wu2018} and NRL-TSI \citep{coddington2016} (0.28 and 0.55 $Wm^{-2}$, respectively), and several $Wm^{-2}$ obtained by models that postulate a long term variation of the quiet Sun, as \citet{egorova2018} or \citet{shapiro2011b}, the latter not shown in Fig2.5 $Wm^{-2}$.~\ref{TSI_rec}.\\ 
Our reconstruction shows a reduced secular modulation compared to estimates by \citet{egorova2018} and \citet{shapiro2011b}. This is due to two important differences with respect to their approach:
$i)$ we modulated a (quiet) network component, whose radiative emission is empirically derived from observations; $ii)$ we took into account only of secular variations of the modulation potential, by empirically decomposing the signal into six components and discarding variations ascribable to the Hale magnetic solar cycle (11-years and 22-years components). The long term modulation has been obtained as the weighted sum of the considered components. The weights have been assigned so that the long-term trend will be present in the modulation of the solar potential but not in the sunspots coverage areas series.\\
Both of these aspects affect the amplitude and shape of the modelled variability. \citet{egorova2018} noted the the amplitude of the variability is strongly affected by the choice of the atmosphere model modulated by the solar modulation potential. The fact that we modulated a bright quiet network component most likely is one of the major factors producing lower TSI variability from the Maunder minimum to modern times than \citet{shapiro2011} and \citet{egorova2018}. The fact that we employed higher order Empirical Modes of the solar modulation potential instead of a 22-years smoothed mean most likely explain the differences found in the shape and phase of the TSI variability. In particular, we note that during the Maunder minimum our reconstructed TSI presents a slow decline and only a slightly faster increase starting around 1680, while reconstructions by \citet{shapiro2011} and \citet{egorova2018} present a steeper decrease, reach a minimum around 1700, and rapidly increase. Similar differences are found for the Dalton minimum and the 1900 minimum.\\ 
We note that our reconstruction for the period before 1751 is closely linked to the trend in the solar modulation potential, which produces an inevitable modulation during the Maunder minimum. There are indications in literature of a residual magnetic activity during that period \citep[e.g.][]{zolotova2015, kopp2016b,jungclaus2017}, smaller than it is reported here. Our result is related to assumption of the constancy of the correlations in Eq.~\ref{Ts_plage_vs_phi_fit} back in time. However, that does not affect the final result of the average trend of the TSI shown in bold red line in Fig.\ref{TSI_rec}.\\
Recently, limits to the variability of the TSI from the Maunder minimum to the present were estimated by the works of \citet{lockwood2020} and \citet{yeo2020}.
\citet{lockwood2020} assumed that the difference between current models and composite measurements provide an estimate of the quiet Sun secular component. By correlating these differences with the solar modulation potential, they estimated that the TSI changes between the Maunder minimum and the present must range between -0.95 and 1.98 $Wm^{-2}$, the negative value indicating that the TSI decreased from the Maunder minimum to the present. \citet{yeo2020} suggested instead that the quiet possible state for the solar photosphere can be represented by local-dynamo three dimensional magneto-hydrodynamic simulations, and estimated an upper limit increase from the Maunder minimum to the present of about 2 $Wm^{-2}$. Our estimate of about 2.5 $Wm^{-2}$ is consistent with these values.\\
It is noteworthy that the TSI exhibits particularly high values during solar cycle 3 (Fig. \Ref{TSI_rec}), although compatible within the error with other TSI reconstructions. As discussed in section \ref{sec:sec5}, the value of the TSI depends primarily on the disk coverage by faculae (plages) and sunspots, and this coverage was significantly high during that particular cycle (see Fig.\Ref{fac_rec} and Fig.\ref{spot_rec}). The analysis of Fig.\Ref{potential_vs_spot} suggests that this is due to the trend in the solar modulation potential $\Phi$, which remained at high levels during cycle 3 and 4. This behavior in $\Phi$ is a consequence of the narrow and deep minimum in \textsuperscript{14}C radiocarbon measurements observed around the maxima of these two cycles.
This deep \textsuperscript{14}C minimum is present in the dataset used in this work, but it is also reported in other datasets \citep[e.g.][]{brehm2021}.
\\
In conclusion, the main findings of our work can be summarized in the following five points:
\begin{enumerate}
    \item we reconstruct the area coverage of faculae (plages) and sunspots from 1513 to 2001 using their cycle-by-cycle observed correlation with the solar modulation potential $\Phi$ over the last century.
    \item we estimate the long-term modulation in the TSI and separate the contributions to the TSI of the different solar magnetic structures through the Empirical Mode Decomposition of the solar modulation potential $\Phi$;
    \item our Total Solar Irradiance reconstructed time series from 1513 to 2001 shows a behavior that is somewhat intermediate between other reconstructions proposed in the literature;
    \item we estimate that the change in TSI levels between the Maunder minimum and the present epoch is approximately $2.5 W m^{-2}$;
    \item we estimate a growth in the TSI value of about $1.5 W m^{-2}$ during the first half of last century. After the 1950s, this value has remained substantially constant (on average) until the beginning of this century.
\end{enumerate}
\acknowledgments
The National Solar Observatory is operated by the Association of Universities for Research in Astronomy, Inc. (AURA), under cooperative agreement with the National Science Foundation. L.B. and S.C. are members of the international team on Reconstructing Solar and
Heliospheric Magnetic Field Evolution Over the Past Century supported by the International Space Science Institute (ISSI), Bern, Switzerland.
This research is partially supported by the Italian MIUR-PRIN \emph{Circumterrestrial Environment: Impact of Sun-Earth Interaction\/} grant 2017APKP7T. This paper is partially based on PhD’s thesis \emph{Magnetic Activity as a driver of the variability of our star\/} conducted and PhD’s thesis \emph{Advanced analysis algorithms for low SNR astrophysical signals\/} conducted by M. Cantoresi and P. Giobbi, respectively, under the supervision of Prof. F. Berrilli. 
M. Cantoresi and P. Giobbi are supported by the Joint Research PhD Program in “Astronomy, Astrophysics and Space Science” between the universities of Roma Tor Vergata and Roma Sapienza, and INAF.

\bibliography{main}{}
\bibliographystyle{aasjournal}

\end{document}